\documentclass[11pt]{amsart}
\allowdisplaybreaks[1]

\usepackage{a4wide}
\usepackage{amssymb}
\usepackage{amsmath}

\newtheorem{theorem}{Theorem}[section]
\newtheorem{lemma}[theorem]{Lemma}
\newtheorem{prop}[theorem]{Proposition}
\newtheorem{coro}[theorem]{Corollary}
\newtheorem{definition}[theorem]{Definition}

\newcommand{\RR}{\mathbb{R}}

\newcommand{\CC}{\mathbb{C}}
\newcommand{\NN}{\mathbb{N}}
\newcommand{\TT}{\mathbb{T}}

\newcommand{\dd}{\,{\rm d}}

\newcommand{\oplam}{\mbox{\Large $\curlywedge$}}

\newcommand{\MM}{\mathcal{M}(G)}

\newcommand{\eltwo}{ \mbox{L}^2 (\varOmega, m) }

\newcommand{\cHp}{\mathcal{H}_{pp}(T)}

\newcommand{\cA}{\mathcal{A}}

\newcommand{\cU}{\mathcal{U}}

\newcommand{\cH}{\mathcal{H}}

\newcommand{\cF}{\mathcal{F} (G)}

\newcommand{\Xil}{\Xi (\varLambda)}
\newcommand{\Xig}{\Xi (G)}
\newcommand{\vL}{\varLambda}

\newcommand{\Lperp}{\widetilde{L}^\perp}
\newcommand{\Lnull}{L^\circ}

\newcommand{\ts}{\hspace{0.5pt}}

\newcommand{\Ghat}{\widehat{G}}
\newcommand{\gammahat}{\widehat{\gamma}}

\DeclareMathOperator{\supp}{supp}

\begin{document}

\title[Point spectrum for measure dynamical systems]
 {Pure Point spectrum for measure dynamical systems on \\ \hspace*{-0.5em}
 locally compact Abelian groups}

\author{Daniel Lenz}
\address{Friedrich-Schiller-Universit\"at Jena, 
          Fakult\"at f\"ur Mathematik und Informatik,
          Mathematisches Institut,
          D-07737 Jena, Germany}
\email{daniel.lenz@uni-jena.de}

\author{Nicolae Strungaru}
\address{Mathematics and Statistics,
University of Victoria, PO BOX 3045 STN CSC Victoria, B.C. Canada
V8W 3P4 {\rm and}
Institute of Mathematics "Simion Stoilow", Bucharest, Romania}
\email{nicolae@math.uvic.ca}

\begin{abstract}
We show equivalence of pure point diffraction and pure point dynamical
spectrum for measurable dynamical systems built from locally finite
measures on locally compact Abelian groups.  This generalizes all
earlier results of this type. Our approach is based on a study of  almost
periodicity in a Hilbert space. It allows us to set up a perturbation
theory for arbitrary equivariant measurable perturbations.
\end{abstract}


\maketitle

\section{Introduction}
This paper deals with mathematical diffraction theory and its
relationship to dynamical systems.  Our main motivation comes from the
study of aperiodic order.

The study of (dis)order is a key issue in mathematics and physics
today. Various regimes of disorder have attracted particular attention
in recent years. A most prominent one is long range aperiodic order
or, for short, aperiodic order.  There is no axiomatic framework
for aperiodic order yet. It is commonly understood to mean a form of
(dis)order at the very border between periodicity and disorder. While
giving a precise meaning to this remains one of the fundamental
mathematical challenges in the field, a wide range of distinctive
feature has been studied in such diverse fields as combinatorics,
discrete geometry, harmonic analysis, K-theory and Schr\"odinger
operators (see e.g. the monographs and proceeding volumes \cite{BMbook,Jan,Moodybook,Pat,Sen,SHS}).

Part of this research is certainly
triggered by the actual discovery of physical substances exhibiting
this form of disorder twenty five  years ago \cite{Shechtman,Ni}.  These
substances were discovered experimentally by their unusual and rather striking
diffraction patterns. These exhibit a (large) pure point component
(meaning order) with symmetries incompatible with periodicity (meaning
aperiodicity). Of course, the discovery of quasicrystals by
diffraction experiments lead to  a particular interest in diffraction
theory of aperiodic order.  Besides this externally motivation, there
also is a strong intrinsic mathematical interest in diffraction theory.

In order to be more precise on this point, let us shortly and with
some grains of salt describe mathematical diffraction theory (see
Section \ref{Diffraction} for details). In mathematical diffraction
theory, the solid in question is modeled by a measure.  The
diffraction is then described by the Fourier transform of this
measure.  The basic intuition is now that order in the original
measure will show up as a (large) pure point component in its Fourier
transform. A particular instance of this intuition is given by the Poisson
summation formula.  To extend this intuition has been a driving
force of the conceptual mathematical study of diffraction for
aperiodic order (see e.g.  Lagarias' article \cite{Lag}).  Let us
emphasize that this conceptual mathematical question has already
attracted attention  before the dawn of quasicrystals, as can be
seen e.g in Meyer's book \cite{Meyer} or the corresponding chapters in
Queffelec's book \cite{Que}.

As mentioned already, we will be concerned with the connection of
diffraction theory and dynamical systems.  Recall that (dis)order
is commonly modeled by dynamical systems.  The elements of the dynamical system then represent the various manifestations of the 'same' form of disorder.  This dynamical system then induces a unitary representation of
the translation group.  The spectrum of the dynamical system is the
spectrum of this unitary representation. Starting with the work of
Dworkin \cite{Dwo}, it has been shown in various degrees of generality
\cite{Hof2,Martin2} that the dynamical spectrum contains the
diffraction spectrum (see \cite{Que} for a similar statement as
well). On the other hand, by the work of van Enter and Mi\c{e}kisz
\cite{EM} it is clear that in general the dynamical spectrum may be
strictly larger than the diffraction spectrum.

In view of the results of \cite{EM}, it is most remarkable that the
two spectra are yet equivalent once it comes to pure point
spectrum. More precisely, pure point dynamical spectrum is equivalent
to pure point diffraction spectrum.  This type of result has been
obtained by various groups in recent years \cite{LMS-1,BL,Gouere-1}.
First Lee/Moody/Solomyak \cite{LMS-1} showed the equivalence for
uniquely ergodic dynamical systems of point sets in Euclidean space
satisfying a strong local regularity condition viz finite local
complexity. Their result was then extended by Gou\'{e}r\'{e}
\cite{Gouere-1} and by Baake/Lenz \cite{BL} to more general
contexts. In particular, it was freed from the assumptions of unique
ergodicity and finite local complexity.  While there is some overlap
between \cite{Gouere-1} and \cite{BL}, these works are quite different
in terms of models and methods.  Gou\'{e}r\'{e} deals with measurable
point processes in Euclidean space, using Palm measures and
Bohr/Besicovich almost periodicity.  Thus, his result is set in the
measurable category.  Baake/Lenz leave the context of points
altogether by dealing with translation bounded measures on locally
compact abelian groups.  Their results are then, however, restricted
to a topological context. In fact, a key step in their setting is to
replace the combinatorial analysis of \cite{LMS-1} by a suitable
application of the Stone/Weierstrass theorem.

Given this state of affairs it is natural to ask whether the
corresponding results of \cite{Gouere-1} and \cite{BL} on dynamical systems can be
unified. This amounts to developing a diffraction theory based on
measure dynamical systems in the measurable category. This is not only
of theoretical  interest. It is also relevant for
perturbation theory. More precisely, one may well argue that aperiodic
order is topological in nature and, hence, a treatment of aperiodic
order in the topological category suffices. However, a more realistic
treatment should allow for perturbations as well. By their very
nature, these perturbations should not be restricted to the
topological category. They should rather be as general as possible. In order to
accommodate this a measurable framework seems highly desirable. The overall aim of this article is then to provide such a framework.  More precisely, the aims are to

\begin{itemize}

\item develop a diffraction theory for measure dynamical systems unifying the
corresponding treatments of \cite{Gouere-1,Gouere-2} and \cite{BL},

\item set up a measurable perturbation theory for these systems.

\end{itemize}

Along our way, we will actually present

\begin{itemize}

\item  a new method of proving the equivalence of pure point diffraction and
pure point dynamical spectrum based on a stability result for the
pure point subspace of a unitary representation.

\end{itemize}

This stability result may be of independent interest.  Its proof is
close in spirit to considerations of \cite{Gouere-1} by relying on
almost periodicity on Hilbert space. It also ties in well with other
recent work focusing on almost periodicity in the study of pure point
diffraction \cite{BM,LR,Str}.

\medskip

The paper is organized as follows:

In Section~\ref{Point}, we discuss some general facts concerning
the point spectrum of a strongly continuous unitary
representation. The main abstract result, Theorem
\ref{stabilitypp}, gives a stability result for the pure point
subspace. This result is then applied to measure dynamical systems
and gives Corollary~\ref{maineins}. This corollary establishes
that the subspace belonging to the point spectrum is invariant
under composition with bounded functions. These results are the
main abstract new ingredients in our reasoning.  They may be
useful in other situations as well. The dynamical systems we are
dealing with are introduced in Section \ref{Measure}. They are
built from locally finite measures on locally compact Abelian
(LCA) groups. We study  a dense set of functions on the
corresponding $L^2$-space and use it to obtain  strong continuity of the associated representation of $G$ in Theorem~\ref{strongly}. The setting for diffraction
theory is discussed in Section~\ref{Diffraction}. As shown there,
the topological approach of \cite{BL} can be extended to a
measurable setting, once a certain finiteness assumption is made.
In particular, there is an abstract way to define the
autocorrelation measure, Prop~\ref{Existenz}. We then come to the
relationship between diffraction and the spectral theory of the
dynamical systems in Section \ref{Dynamical}.  The crucial link is
provided by Theorem~\ref{Spectralmeasure} which states that the
Fourier transform of the autocorrelation is a spectral measure for
a subrepresentation.  When combined with the abstract results of
Section \ref{Point}, this gives Theorem~\ref{main} showing the
equivalence of the two notions of pure point spectrum. In Section
\ref{Perturbation} we use our results to briefly  set up a perturbation
theory.

\section{Point spectrum of strongly continuous unitary representations and measurable dynamical systems} \label{Point}

In this section, we discuss the pure point subspace of a strongly
continuous unitary representation. We obtain an abstract stability
result for this subspace and apply it to   dynamical systems.

\medskip

Let $G$ be a locally compact, $\sigma$-compact, Abelian group. The
dual group of $G$ is denoted by $\widehat{G}$, and the pairing between
a character $\lambda\in \widehat{G}$ and an element $t\in G$ is
written as $(\lambda, t)$, which, of course, is a number on the unit circle.

A unitary representation $T$ of $G$ in the Hilbert space $\cH$ is a
group homomorphism into the group of unitary operators on $\cH$. It is
called strongly continuous if the map $ G\longrightarrow \cH$,
$t\mapsto T^t f$, is continuous for each $f\in \cH$. As usual, the
inner product on a Hilbert space is denoted by $\langle \cdot,\cdot
\rangle$.

A non-zero $f\in\cH$ is called an
\textit{eigenfunction} of $T$ if there exists a
$\lambda\in \widehat{G}$ with $T^t f = (\lambda, t) f$ for every $t\in
G$. The closure  of the linear span of all
eigenfunctions of $T$ will be denoted by $\cHp$. $T$ is said to have
\textit{pure point spectrum}, if $\cHp = \cH$.

\smallskip

For a strongly continuous $T$  there exists by
Stone's Theorem (compare \cite{Loomis}) a map
\begin{equation*}
   E_T\!:\, \mbox{Borel sets on $\widehat{G}$}\; \longrightarrow \;
   \mbox{Projections on $\cH$}
\end{equation*}
with
\begin{itemize}
\item $E_T (\emptyset) = 0$, $E_T (G) = Identity$,

\item $E_T (A) = \oplus E_T (A_j)$ whenever $A$ is the disjoint union of the Borel sets $A_j$, $j\in \NN$,

\end{itemize}

such that
\begin{equation*}
   \langle f, T^t f \rangle = \int_{\widehat{G}} (\lambda, t)\dd
   \langle f, E_T(\lambda) f \rangle =: \int_{\widehat{G}} (\lambda,
   t)\dd  \rho^{}_f (\lambda),
\end{equation*}
where $\rho^{}_f$ is the measure on $\widehat{G}$ defined by
$\rho^{}_f (B) := \langle f, E_T (B)f\rangle$. Because of its properties $E_T$ is  called a projection valued measure.

A continuous function $p$ on $G$ with values in a Banach
space $(X,\|\cdot\|)$ (e.g. $X=\CC$ or $X=\cH$) is called \textit{almost
periodic} if for every $\varepsilon >0$, the set
$$\{t\in G : \|p( t +s) - p(s)\| <\varepsilon \;\:\mbox{for all $s\in
G$}\}$$ is relatively dense in $G$.  Here, a subset $S$ of $G$ is called relatively dense if there exists a compact subset $K$ of $G$ with $S+ K = G$.  If $p(t)=T^t x$ for some $x
\in \cH$, and a strongly continuous  $T$ then the almost periodicity of $p$ is equivalent to the
closure $\overline{p(G)}$ of $p(G)$ being compact.

\smallskip

We can now formulate the following characterization of $\cHp$.

\begin{lemma}\label{characterizationpp} Let $T$ be a strongly continuous unitary representation of $G$ on $\cH$. Then, the following assertions are equivalent for $f\in \cH$:

\begin{itemize}

\item[(i)] The map $G\longrightarrow \cH$, $t\mapsto T^t f$, is almost periodic.

\item[(ii)] The  map $G\longrightarrow \CC$, $t\mapsto \langle f, T^t f \rangle $, is almost periodic.

\item[(iii)] $\rho_f$ is a pure point measure.

\item[(iv)]  $f$ belongs to $\cHp$.
\end{itemize}
\end{lemma}

\begin{proof}
The equivalence of $(iii)$ and $(iv)$ is standard. The equivalence of
$(ii)$ and $(iii)$ follows by a result of Wiener as $t \mapsto \langle
f, T^t f\rangle$ is the Fourier transform of $\rho_f$.  The
implication $(i)\Longrightarrow (ii)$ is clear.
It remains to show  $(ii)\Longrightarrow (i)$: A direct calculation gives
$$\|f - T^t f\|^2 = 2 \langle f, f\rangle - \langle f, T^t f\rangle -
\overline{\langle f, T^t f\rangle} \leq 2 | \langle f, f \rangle -
\langle f, T^t f\rangle |.$$ Now, the desired result follows.
\end{proof}

For our further analysis, we need some more pieces of notation. A measure $\rho$
on $\Ghat$ is said to be supported on the  subset $S$ of $\Ghat$ if there exists a measurable subset $S'$ of $S$ with
$\rho (\Ghat\setminus S') =0$. For a subgroup $S$ of $\Ghat$ equipped
with the discrete topology (which may not be the topology induced by $G$!), the dual group $\widehat{S}$ is compact. The
injective group homomorphism ${S} \longrightarrow \Ghat$,
$\lambda\mapsto \lambda$, induces the  group homomorphism $G\longrightarrow \widehat{S}$, $t\mapsto (
\lambda \mapsto (\lambda,t))$. The latter will be  denoted by $j$. It has
dense range. Conversely, if $\TT$ is a compact group and
$j: G\longrightarrow \TT$ is a continuous
group homomorphism with dense range, then $\widehat{\TT}$ can naturally be
considered to be a subgroup of $\Ghat$ with the discrete topology via $i
:\widehat{\TT} \longrightarrow \Ghat$, $i(\lambda) (t) :=(\lambda,j(t))$.

\begin{lemma}\label{mainsupportinglemma}
Let  a strongly continuous unitary representation $T$  of $G$ on
  $\cH$ and $f\in \cH$ be given.

(a) If $f$ belongs to $\cHp$ with $\rho_f$ supported on the subgroup $S$ of
$\Ghat$ and  $j: G\longrightarrow \widehat{S}$ is the canonical group
homomorphism, then  $t\mapsto T^t f$ can be lifted to a continuous
map on $\widehat{S}$, i.e. there exists a continuous map $u :
\widehat{S}\longrightarrow \cH$ with $u\circ j (t) = T^t f$ for every $t\in
G$.

\smallskip

(b) Let $\TT$ be a compact group and $j: G\longrightarrow \TT$ a continuous
group homomorphism with dense range. If $u : \TT\longrightarrow \cH$ is
continuous with $u\circ j (t) = T^t f$ for every $t\in G$, then $f$ belongs to
$\cHp$ and $\rho_f$ is supported on $i(\widehat{\TT})$.
\end{lemma}
\begin{proof}
(a) As $f$ belongs to $\cHp$ and  $\rho_f$ is supported on $S$, we have $f = \sum_{\lambda\in S}
c_\lambda f_\lambda$ with  $f_\lambda$ which are either $0$ or normalized eigenfunctions  to
$\lambda$. Then, $\sum_{\lambda \in S} |c_\lambda|^2< \infty$ as the $f_\lambda$ are pairwise orthogonal. As   $|(\sigma,\lambda)|$ has  modulus one for each $\lambda\in S$ and $\sigma \in \widehat{S}$ and $\sigma \mapsto (\sigma,\lambda)$ is continuous,  the map, $ u: \widehat{S} \longrightarrow \cH$,
$$u(\sigma) :=\sum_{\lambda\in S} (\sigma,\lambda) c_\lambda f_\lambda$$
can easily be seen to have  the desired properties.

\smallskip

 (b) By $u\circ j (t) = T^t f$, we infer that $\{T^t f: t\in G\}$ is
contained in $u(\TT)$, which is compact as $u$ is continuous and $\TT$
is compact. Hence, $f$ belongs to $\cHp$. Moreover, by
$\widehat{\rho_f} (t) = \langle f, T^t f\rangle = \langle f, u\circ
j(t)\rangle$, we infer that $\widehat{\rho_f} (t)$ can be lifted to a
continuous function $g$ on $\TT$. Note that $g$ is positive definite
as $\widehat{\rho_f} (t)$ is positive definite and $j$ has dense range.  Since $\TT$ is a
compact group and $g$ is positive definite and continuous, we can write
$$g= \sum_{k=1}^N a_k \chi_k .$$
Here, $N=\infty$ or  $N\in \NN$ and the $\chi_k$ belong to
$\widehat{ \TT}$ and the sum on the right converges uniformly.
Evaluating $g$ on $j(t)$ for some $t\in G$ we then obtain
$$ \widehat{\rho_f} (t)  =\sum_{k=1}^N a_k \chi_k \circ j  (t)\,.$$
Since the  (inverse) Fourier transform is continuous in the uniform topology, we obtain by taking the  inverse Fourier transform
$$\rho_f =  \sum_{k=1}^N a_k \delta_{- i (\chi_k)}  \,,$$
with the sum on the right converging in the vague topology.
This gives that $\rho_f$ is supported on $i(\widehat{\TT})\subset
\Ghat$.
\end{proof}

These lemmas yield the following abstract stability result for $\cHp$.

\begin{theorem}\label{stabilitypp}
Let $T$ be a strongly continuous unitary representation of $G$ on
$\cH$. Let $C : \cH \longrightarrow \cH$ be continuous with $T^t C f =
C T^t f$ for each $t\in G$ and $f\in \cH$.  Then, $C$ maps $\cHp$ into
$\cHp$. If $f$ belongs to $\cHp$ and $\rho_f$ is supported on the subgroup $S$
of $\Ghat$, then so is $\rho_{Cf}$.
\end{theorem}

\textbf{Remark.} Let us emphasize  that $C$ is not assumed to be linear.

\begin{proof}
  Choose $f\in \cHp$ arbitrary.  Let $A:=\overline{\{T^t C f: t\in G\}}$ and
  $B:=\overline{\{ T^t f : t\in G\}}$. Then, $B$ is compact by the Lemma
  \ref{characterizationpp} As $C$ is continuous and commutes with $T$ this
  yields that $A = C(B)$ is compact as well.  Then, by Lemma
  \ref{characterizationpp} again, $Cf$ belongs to $\cHp$.

It remains to show the statement about $\rho_{Cf}$: Equip $S$ with the
discrete topology and denote its compact dual group by $\TT$. As $\rho_f$ is
supported in $S$, part (a) of the previous lemma shows that $t\mapsto T^t f$
can be lifted to a  continuous map $u$ on $\TT$. As $C$ is continuous,
$t\mapsto C T^t f = T^t C f$, then lifts to the continuous map $ C\circ u :
\TT \longrightarrow \cH$. By (b) of the previous lemma, $\rho_{Cf}$ is then
supported in $S$ as well.
\end{proof}

We  now come to an application of these considerations to
measurable dynamical systems.
Let a measurable space
$(\varOmega,\varSigma_\varOmega)$ consisting of a set $\varOmega$ and
a $\sigma$-algebra $\varSigma_\varOmega$ on it be given. Let
\begin{equation*} \label{def:action}
  \alpha \!:\, G \times \varOmega \;\longrightarrow\; \varOmega
\end{equation*}
be an action which is  measurable in each variable. Then $(\varOmega,
\alpha)$ is called a measurable dynamical system.
Let $m$ be a $G$-invariant probability measure on $\varOmega$ and
denote the set of square integrable functions on $\varOmega$, with
respect to $m$, by $L^2 (\varOmega,m)$. This space is equipped with
the inner product $\langle f, g\rangle := \int \overline{f (\omega)}
g(\omega)\dd m(\omega)$. The action $\alpha$ induces a unitary
representation $T$ of $G$ on $L^2 (\varOmega,m)$ in the obvious way,
namely $T^t h$ is given by $(T^t h)(\omega) := h(\alpha^{}_{-t}
(\omega))$.

Let $C_c (\CC)$ be  the set of continuous functions on $\CC$
with compact support. Then, the following holds.

\begin{lemma}
Let $(\varOmega,\Sigma_\varOmega)$ be a measure space with a
probability measure $m$.  For each $g\in C_c (\CC)$, the map $C_g :
\eltwo\longrightarrow \eltwo$, $f\mapsto g\circ f$, is uniformly continuous.
\end{lemma}
\begin{proof}
 Choose $\varepsilon >0$ arbitrary. As $g$ is uniformly continuous, there exists a $\delta>0$ such that
$$ | g(x) - g(y) |^2 \leq \frac{\varepsilon}{2}
\,\:\mbox{whenever}\;\: |x - y| \leq \delta.$$ Set $M:=\max\{
|g(x)| : x \in \CC\}$. By a direct Tchebycheff type estimate we have
for arbitrary $h,h'\in \eltwo$
$$ m ( \varOmega_{\delta,h,h'} ) \leq \frac{\|h - h '\|^2}{ \delta^2}$$
where
$$ \varOmega_{\delta,h,h'}  = \{ \omega\in \varOmega : |h (\omega) - h'
(\omega)| \geq \delta\}.$$
Setting $D_{h,h'} := |g\circ h - g\circ h'|$ we obtain
\begin{eqnarray*}
\int_\varOmega D_{h,h'}^2 d m (\omega) &=&
\int_{\varOmega_{\delta,h,h'}} D_{h,h'}^2 d m +
\int_{\varOmega\setminus \varOmega_{\delta,h,h'}} D_{h,h'}^2 d m \\
&\leq & m(\varOmega_{\delta,h,h'}) \, 4 M^2 + m ( \varOmega\setminus \varOmega_{\delta,h,h'} ) \, \frac{\varepsilon}{2} \\
& \leq &  \frac{ 4 M^2 \|h - h'\|^2}{\delta^2} + \frac{\varepsilon}{2}.
\end{eqnarray*}
This finishes the proof.
\end{proof}

\begin{coro} \label{maineins}
  Let\/ $(\varOmega, \alpha)$ be a measurable dynamical system and\/
  $m$ an\/ $\alpha$-invariant probability measure on\/ $\varOmega$
  such that the associated unitary representation is strongly
  continuous.  Then, for arbitrary\/ $f\in \cHp$ and\/ $g\in C_c
  (\CC)$, the function\/ $g\circ f$ belongs to\/ $\cHp$ and if  $\rho_f$ is
  supported on the subgroup $S$ of $\Ghat$, so is $\rho_{g\circ f}$.
\end{coro}
\begin{proof}
This follows from the previous lemma and Theorem \ref{stabilitypp}.
\end{proof}

We also note the following  result on compatibility of almost periodicity with products, which is a slight generalization of Lemma $1$ of \cite{BL} and Lemma $3.7$ in  \cite{LMS-1}.

\begin{lemma} \label{multiplication}
  Let\/ $(\varOmega, \alpha)$ be a measurable dynamical system and\/
  $m$ an\/ $\alpha$-invariant probability measure on\/ $\varOmega$
  such that the associated unitary representation is strongly
  continuous.  Let  $f$ and $g$ be bounded functions in  $\cHp$ such that  $\rho_f$ and $\rho_g$ are supported on the subgroup $S$ of $\Ghat$. Then, $f g$ is a bounded function in $\cHp$ and  $\rho_{f g}$ is supported on $S$ as well.
  \end{lemma}
\begin{proof}
It is shown in Lemma $1$ of  \cite{BL} that the product of bounded functions in $\cHp$ belongs again to $\cHp$. Here, we give a different proof, which shows the statement on the support of the spectral measures as well.
By  (a) of Lemma \ref{mainsupportinglemma}, there exist continuous maps $u_f : \widehat{S}\longrightarrow L^2 (\varOmega,m)$ and $u_g : \widehat{S}\longrightarrow L^2 (\varOmega,m)$ with $u_f\circ j (t) =T^t f$ and $u_g \circ j (t) = T^t g$ for all $t\in G$. Then,  using the boundedness of $f$ and $g$, we can easily infer that
$$u:= u_f u_g  : \widetilde{S}\longrightarrow L^2 (\varOmega,m)$$
 is continuous. By construction we have  $u\circ j(t) = T^t (f g)$ for all $t\in G$. Thus, the desired statement follows from (b) of Lemma \ref{mainsupportinglemma}.
\end{proof}

\section{Measure dynamical systems}\label{Measure}
In this section, we introduce the measurable dynamical systems we are
dealing with. These will be dynamical systems of measures on
groups. These systems are interesting objects in their own right.
Moreover, as discussed in the next section, they provide an
adequate framework for diffraction theory.

\medskip

Let $G$ be the fixed $\sigma$-compact LCA group. The set of
 continuous functions on $G$ with compact support is denoted by $C_c
 (G)$.  It is equipped with the locally convex limit topology induced
 by the canonical embeddings $C_K (G)\hookrightarrow C_c (G)$, where
 $C_K (G)$ is the space of complex continuous functions with support
 in $K\subset G$ compact. The support of $\varphi\in C_c(G)$ is
 denoted by $\supp (\varphi)$.  The set $\MM$ is then defined to be the dual of the
 the space of $C_c (G)$ i.e. the space of continuous linear functionals on $C_c(G)$. The elements of $\MM$ can be considered as complex measures. The total variation $|\mu|$ of an element of $\MM$ is again an element of $\MM$ and in fact a positive regular Borel measure characterized by
 $$|\mu| (\varphi) =\sup\{|\mu(\psi)| : \psi \in C_c (G)\;\mbox{real valued with $|\psi|\leq \varphi$}\}$$
 for every nonnegative $\varphi \in C_c (G)$. Moreover, there exist a measurable $u: G\longrightarrow \CC$
with $|u(t)|=1$ for $|\mu|$-almost every $t\in G$  with
$$ \mu (\varphi) = \int u \varphi d|\mu|$$
for every $\varphi \in C_c (G)$. This allows us in particular to define the restriction of $\mu$ to subsets of $G$ in the obvious way.

 The
 space $\MM$ carries the vague topology. This topology equals the
 weak-$\ast$ topology of $C_c(G)^\ast$, i.e., it is the weakest
 topology which makes all functionals $\mu\mapsto \mu(\varphi)$,
 $\varphi\in C_c (G)$, continuous. Thus, if we define
\begin{equation*}
 f\! :\, C_c (G) \; \longrightarrow \;
    \{\mbox{functions on $\MM$}\}\, , \;
    \varphi  \mapsto f_\varphi\, ,\;\: \mbox{by}\;\:
   f_\varphi (\mu):= \int_G \varphi (-s) \dd\mu(s),
\end{equation*}
then the topology is generated by
\begin{equation*}
   \{f_\varphi^{-1} (O) : \varphi\in C_c (G),\,  O\subset\CC\mbox{ open}\}.
\end{equation*}
Here, the reader might wonder about the sign in the definition of
$f_\varphi$. This sign is not necessary. However, it  does not
matter either as $G\longrightarrow G$, $s\mapsto s^{-1}$, is a
homeomorphism. It will simplify some formulae later on.

We will be concerned with measurable dynamical systems consisting of
elements of $\MM$. Thus, we need a $\sigma$-algebra on $\MM$ and an
action of $G$. These will be provided next. We start with the
$\sigma$-algebra.  As discussed above, $\MM$ is a topological
space. Thus, it carries a natural $\sigma$-algebra, namely the Borel
$\sigma$-algebra generated by the open sets. Denote this algebra by
$\varSigma_{\MM}$.

\medskip

\textbf{Remark.} If $G$ has a countable basis of the topology then the restriction of the Borel $\sigma$-algebra  to the set $\MM_+$ of
nonnegative measures is the $\sigma$-algebra $\Sigma'$  generated by
$\{ f_\varphi^{-1} (O)\cap \MM_+ : \varphi \in C_c (G), O\subset \CC\mbox{ open}\}$. This is a consequence of the well known second countability
of the vague topology on $\MM_+$ (See Chapter IV, Section 31 in \cite{Bauer}). A proof can be given along the following line:
The set $\MM_+$ with the vague topology is a second countable metric space and a metric can be given as
$$ d(\mu,\nu):=\sum_{n\in \NN} \frac{ |f_{\varphi_n}(\mu) - f_{\varphi_n} (\nu)|}{ 2^n (1 + |f_{\varphi_n}(\mu) - f_{\varphi_n} (\nu)|)}$$
with a suitable dense set $\{\varphi_n : n\in \NN\}$ in $C_c (G)$. The definition of the metric shows that all balls $B_s(\mu):=\{\nu \in \MM_+ : d(\mu,\nu) < s\}$, $\mu\in \MM_+$, $s\geq 0$, belong to $\Sigma'$. This is then true for countable unions of such balls as well and the statement follows.

\medskip

\begin{lemma}
The map $f_\varphi \circ | \cdot|$ is measurable for every $\varphi
\in C_c (G)$. In particular, the map $\MM\longrightarrow \MM$, $\mu
\mapsto |\mu|$, is measurable.
\end{lemma}
\begin{proof}
It suffices to show that $f_\varphi \circ | \cdot|$ is measurable for
every $\varphi \in C_c (G)$ with $\varphi \geq 0$. Standard theory
(see Chapter 6 in \cite{Ped} and Proposition $1$ in \cite{BL}) gives
$$ f_\varphi (|\mu|) = \sup\{ | f_{\psi \varphi} (\mu)| : \psi \in C_c
(G), \|\psi\|_\infty\leq 1\}.$$ As $\mu \mapsto |f_{\psi \varphi}
(\mu)|$ is continuous,  $f_\varphi \circ | \cdot|$ is
then semicontinuous  and hence measurable.
\end{proof}

As for the action of $G$ on $\MM$, there is a natural action of $G$ on
$\MM$ given by
\begin{equation*}
  \alpha\! :\, G\times \MM \; \longrightarrow \; \MM\, , \;
  \alpha^{}_t (\mu) := \delta^{}_t \ast \mu,
\end{equation*}
where $\delta_t$ is the unit point measure at $t \in G$.
Here, the convolution
$\mu\ast\nu$ of two convolvable elements of $\MM$ is the measure
defined by $\big(\mu\ast\nu\big) (\varphi) := \int_{G\times G}\,
\varphi(s+t) \dd\mu(t) \dd\nu(s)$.

\smallskip

The map $\alpha$ is measurable in each variable, as shown in the next lemma.

\begin{lemma}\label{stetig} (a) For fixed $\mu\in \MM$, the map  \/ $ G \longrightarrow \MM$, $t\mapsto \alpha_t \mu $,
  is continuous, hence also measurable.

(b) For fixed $t\in G$, the map $ \MM \longrightarrow \MM$, $\mu
  \mapsto \alpha_t \mu $, is continuous, hence also measurable.
\end{lemma}
\begin{proof} This is straightforward.
\end{proof}

Putting this together, we see that $\MM$ equipped with the Borel
$\sigma$-algebra and the natural action of $G$ by shifts is a
measurable dynamical system.

As discussed in Section~\ref{Point}, every $\alpha$-invariant
probability measure $m$ on $\MM$ induces a unitary representation $T$
of $G$ on $L^2 ( \MM,m)$.

For further  understanding of this unitary representation, it will be crucial to control it by a suitable set of functions. This set of functions is introduced next.


\begin{definition}\label{dense} Consider the algebra generated by the set
\begin{equation*}\{ g\circ f_\varphi : g\in C_c (\CC),
  \varphi  \in C_c (G)\}.
\end{equation*}
Let $\cA(G)$ be the closure  of this algebra  in the algebra of all continuous bounded functions on $\MM$ equipped with the supremum norm.
An \/ $\alpha$-invariant probability
  measure $m$  on\/  $\MM$ is said to satisfy the denseness assumption (D) if  the algebra $\cA (G)$ is dense  in \/ $L^2 (\MM,m)$.
\end{definition}
Note that condition (D) means that the set of finite products of functions of the form $g\circ f_\varphi$, $g\in C_c (\CC)$, $\varphi \in C_c (G)$, is total in $L^2 (\MM,m)$.

\medskip

We now discuss two instances in which condition (D) holds.

\begin{prop}  Let $m$ be an $\alpha$-invariant probability measure on $\MM$.
If the restriction of the  $\sigma$-algebra of $\MM$ to the support of $m$ is generated by the set $\{f_\varphi^{-1} (O) : \varphi\in C_c (G), O  \subset \CC\mbox{ open}\}$, then (D) holds. In particular, (D) holds whenever $G$ has a countable basis of topology and $m$ is supported on the set of nonnegative measures.
\end{prop}
\begin{proof} As $\CC$ has a countable basis of the topology, the $\sigma$-algebra on $\MM$ is
then generated by the set $\{f_\varphi^{-1} (K) : \varphi\in C_c (G), K
 \subset \CC\mbox{ compact}\}$. In particular, the corresponding set of products of
of characteristic functions
\begin{equation*}
\{ 1_K \circ f_\varphi :\varphi\in C_c (G), K \subset
 \CC\mbox{ compact}\}
\end{equation*}
is total in $L^2 (\MM,m)$. Here, $1_S$ denotes the characteristic
function of $S$. Therefore, it suffices to show that all functions
of the form $1_K \circ f_\varphi, \varphi\in C_c (G)$, $K \subset
\CC$ compact,  can be approximated by functions of the form
$g\circ f_\varphi$ with $g\in C_c (\CC)$. This can be done by
choosing,  for $K\subset \CC$ compact, a compact $L\subset \CC$
containing $K$ and a sequence $(g_n)$ of nonnegative functions in
$C_c (\CC)$ such that $g_n$ converge pointwise to $1_K$, are all
supported in $L$ and are uniformly bounded by, say, $1$.

The 'in particular' statement now follows from the last remark.
\end{proof}

\begin{prop} The condition (D) is satisfied whenever $m$ is supported on a compact $\alpha$-invariant subset of $\MM$.
\end{prop}
\begin{proof} This follows by a Stone/Weierstrass type argument (see \cite{BL} as well): The algebra  in question separate the points, does not vanish identically anywhere  and is closed under complex conjugation. The algebra is then dense in the set of continuous functions on the compact support of $m$. Hence, it is dense in $L^2 (\MM,m)$ as well.
\end{proof}

\textbf{Remark.} The previous propositions imply that (D) holds in all the  settings considered for diffraction so far. More precisely, the setting of uniformly discrete point sets discussed e.g. in the survey article \cite{Lag} its generalization to translation bounded measures \cite{BL} deal with compact subsets of $\MM$. On the other hand
the point process setting first introduced  by  \cite{Gouere-1} deals with $\RR^d$ and hence  admits  a countable basis of topology.


\medskip

\begin{theorem}\label{strongly}
  Let\/ $m$ be an\/ $\alpha$-invariant probability
  measure on\/  $\MM$, which satisfies (D). Then, the representation\/ $T$ is strongly
  continuous.
\end{theorem}
\begin{proof}
As $T^t$ is unitary for every $t\in G$, it is bounded with norm
$1$ uniformly in $t\in G$. By (D), it therefore
suffices to show continuity of $t\mapsto T^t f$ for $f$ a finite product of functions of the form $ g\circ
f_\varphi$ with $\varphi \in C_c(G)$ and $g\in C_c (\CC)$. It suffices to show continuity at $t=0$. As
$$T^t (f_1 f_2) - f_1 f_2 = (T^t f_1 ) T^t f_2 - f_1 f_2 = (T^t f_1) (T^t f_2 - f_2) + (T^t f_1 - f_1) f_2$$
and functions of the form $g\circ f_\varphi$ with $g\in C_c (\CC)$ and $\varphi \in C_c (G)$ are bounded, it suffices to consider  $f= g\circ f_\varphi$.  Let $K$ be
an arbitrary compact neighborhood of $0\in G$.  Let $L$ be a
compact set in $G$ with $L \supset K- \supp \varphi$ and $\psi\in
C_c (G)$ nonnegative with $\psi \equiv 1$ on $L$. Then,
$$ \left| \varphi (t -s) -\varphi (-s) \right|\leq \psi (-s) \|\varphi (t -\cdot)
-\varphi (-\cdot)\|_\infty
$$
for every $s\in G$ and $t\in K$ and, in particular,
\begin{equation*}
(*) \;\:\;\:\;\:\;| f_\varphi (\alpha_{-t} \mu) - f_\varphi (\mu)| \leq \int_G |\varphi (t -s)
 -\varphi (-s)| d|\mu| (s)\leq \|\varphi (t -\cdot) -\varphi
 (-\cdot)\|_\infty f_\psi (|\mu|)
\end{equation*}
for every $t\in K$.
As $\mu\mapsto f_\psi (|\mu|)$ is measurable, the set
$$\varOmega_N :=\{ \mu\in \MM : f_\psi (|\mu|)\leq N\}$$ is measurable
for each $N\in \NN$. Obviously, these sets are increasing and cover
$\MM$. Thus, for each $\varepsilon>0$, there exists $N(\varepsilon)\in
\NN$ with $m(\MM \setminus \varOmega_{N(\varepsilon)}) \leq
\varepsilon$.  Invoking $(*)$, we then get
\begin{eqnarray*}
\|T^t g\circ f_\varphi - g\circ f_\varphi\| ^2 & = & \int_{\MM} |g \circ
f_\varphi (\alpha_{-t} \mu) -g\circ f_\varphi (\mu)|^2 dm(\mu)\\ &= &
\int_{\varOmega_{N(\varepsilon)}} | \cdot|^2 d m (\mu) +
\int_{\MM\setminus \varOmega_{N(\varepsilon)}} | \cdot|^2 d m (\mu) \\
& \leq & B(t,\varepsilon) + \varepsilon 4 \|g\|_\infty^2
\end{eqnarray*}
with
$$ B(t,\varepsilon):= \sup\{ |g(x) - g(y)|^2 : |x -y|\leq
N(\varepsilon) \|\varphi (t - \cdot) - \varphi (-\cdot)\|_\infty \}.$$
As $g$ is uniformly continuous, $B(t,\varepsilon)$ becomes arbitrarily
small for $t$ close to $0$ and $\varepsilon$ fixed. This easily shows
the desired continuity.
\end{proof}

For our further consideration, we will need a certain finiteness
assumption on the probability measure $m$. This assumption is well
known in the theory of stochastic processes. It is given in the next
definition.

\begin{definition}
  The\/ $\alpha$-invariant probability measure\/ $m$ on\/ $\MM$ is
  called square integrable if\/ $f_\varphi \circ |\cdot | :
  \MM\longrightarrow \CC$ belongs to\/ $L^2 ( \MM,m)$ for every\/
  $\varphi\in C_c (G)$.
\end{definition}

\begin{lemma}
Let $m$ be an $\alpha$-invariant square integrable probability
measure\/ $m$ on\/ $\MM$. Then, the map $C_c (G) \longrightarrow L^2
(\MM,m)$, $\varphi \mapsto f_\varphi$, is continuous.
\end{lemma}
\begin{proof}
We have to show that $C_K (G)\longrightarrow L^2 (\MM,m)$, $\varphi
 \mapsto f_\varphi$, is continuous for every compact $K$ in $G$. Let
 $\psi\geq 0$ be a function in $C_c (G)$ with $\psi \equiv 1 $ on
 $K$.  Then,
$$|\varphi (s) | \leq \psi (s) \|\varphi\|_\infty$$
for every $\varphi \in C_K (G)$.  This gives
\begin{eqnarray*}
\| f_\varphi\|^2 = \int \left|\int \varphi(-s) d\omega (s) \right|^2 d
m(\omega) &\leq & \int \left|\int \psi (-s) \|\varphi \|_\infty d
|\omega| (s) \right|^2 d m (\omega)\\ & =& \|\varphi\|_\infty^2 \int
|f_\psi (|\omega|)|^2 d m(\omega)
\end{eqnarray*}
and the desired continuity follows.
\end{proof}

\textbf{Remark.} Note that the lemma gives another proof for the
strong continuity of $T$ in the case of square integrable measures
$m$ satisfying (D). More precisely, for $\varphi \in C_c (G)$ and $g\in C_c (\CC)$, the map $G\longrightarrow  L^2 (\MM,m)$,   $t\mapsto T^t (g\circ f_\varphi)$ can be composed
into the continuous maps $G\longrightarrow C_c (G)$,   $t\mapsto \varphi (\cdot -t)$,  $f : C_c (G)\longrightarrow L^2 (\MM,m)$, and $ C_g: L^2 (\MM,m)\longrightarrow L^2 (\MM,m),    h\mapsto g \circ h$.

\medskip

\begin{definition} Let $m$ be square integrable. Then, \/ $\cU$ is defined to be the  closure of\/ $\{f_\varphi : \varphi \in C_c (G)\}$
  in\/ $L^2 (\MM,m)$.
\end{definition}

\begin{lemma} Let $m$ be square integrable. Then,
   $\cU$ is a\/ $T$-invariant subspace.
\end{lemma}
\begin{proof}
This is immediate from $T^t f_\varphi = f_{\varphi_t}$ with $\varphi_t (s) =
\varphi (t - s)$.
\end{proof}

\section{Diffraction theory and the autocorrelation measure}\label{Diffraction}
In this section, we present a basic setup for diffraction theory for
our measure dynamical systems.  Before we actually start with the
mathematical formulation, we shortly discuss the physical context of
our setting and the relationship of the presented material to earlier work.

In the simplest models for diffraction of a solid, the solid in
question is modeled by a subset $\vL$ of Euclidean space, which
describes the positions of the atoms of the solid. The diffraction of
an incoming beam is then governed by interference of beams scattered
by different points of the solid. Thus, the relevant set is the set of
differences between points of $\vL$ or rather differences averaged
according to occurrence. This yields the so-called autocorrelation
measure $\gamma_\vL$ of $\vL$ given by
$$\gamma_\vL:=\lim_{n\to \infty} \frac{1}{|B_n|} \sum_{x,y\in B_n\cap\vL}
\delta_{x-y} = \lim_{n\to \infty} \frac{1}{|B_n|} \delta_{\vL\cap
B_n} \ast \delta_{-\vL\cap B_n}.$$ Here, $B_n$ is the ball around
the origin with radius $n$, $|B_n|$ is its volume, $\delta_x$
denotes the point measure at $x$ and the limit is assumed to
exist. The diffraction of $\vL$ is then described by the Fourier
transform of $\gamma_\vL$ (see e.g. \cite{Hof} for further details on  this approach).  In order to obtain existence of the
limit, one usually introduces a framework of dynamical systems and
uses an ergodic theorem.

In fact, as shown recently \cite{Gouere-1, Gouere-2} (see
\cite{BL} as well), it is possible to express the limit by a closed
formula. This opens up the possibility to define the autocorrelation by
this closed formula irrespective whether the dynamical system is
ergodic or not. This approach has been taken in \cite{BL}.  In fact,
as argued in \cite{BL, BL2}, it is more appropriate to work
with measures than with point sets. This lead to the notion of measure
dynamical system. In the framework of aperiodic order, it is natural
to restrict attention to topological dynamical systems and this is what
has been analyzed in \cite{BL,BL2}. However, as discussed in the
introduction both from the mathematical point of view and from the
point of view of perturbation theory, it is natural to leave the
topological category and develop diffraction theory in the measurable
category. This is done next.  While our overall line of reasoning
certainly owes to \cite{BL}, we have to overcome various technical
issues. More precisely, as \cite{BL} deals with a topological
situation and compact spaces, all functions $f_\varphi$ (defined
above) were uniformly bounded there. This is not the case here
anymore. To remedy this, we use the assumption of square
integrability.

\smallskip

We start by introducing some further notation: For a measure $\mu$ on
$G$ and a set $B\subset G$, we denote by $\mu_B$ the restriction of
$\mu$ to $B$. For a function $\zeta$ on $G$ we define
$\widetilde{\zeta}$ by $\widetilde{\zeta} (s) :=
\overline{\zeta(-s)}$. For a measure $\mu$ on $G$ we define the
measure $\widetilde{\mu}$ by $\widetilde{\mu} (\varphi) :=
\overline{\mu( \widetilde{\varphi})}$.

\begin{lemma} \label{Existenz}
  Let $m$ be an   $\alpha$-invariant square integrable probability
measure\/ $m$ on\/ $\MM$.
  Let a function\/ $\sigma\in C_c (G)$ be given
  with\/ $\int_G \sigma(t)\dd t = 1$.
  For\/ $\varphi\in C_c (G)$, define
\begin{equation*}
    \gamma^{}_{\sigma,m} (\varphi) \; := \;
    \int_\varOmega \int_G \overline{\int_G \overline{\varphi (s-t) \sigma(t)} \dd \omega(s)} \dd \omega(t)  \dd m(\omega).
\end{equation*}
  Then, the following holds:
\begin{itemize}
\item[\rm (a)] The map\/ $\gamma^{}_{\sigma,m} \!:\, C_c
   (G)\longrightarrow \CC$ is continuous, i.e.,
   $\gamma^{}_{\sigma,m}\in \MM$.
\item[\rm (b)] For\/ $\varphi,\psi \in C_c (G)$, the equation\/
   $\big(\widetilde{\varphi}\ast \psi \ast \gamma^{}_{\sigma,m}\big) (t)
   =  \langle f_\varphi, T^t  f_\psi \rangle$ holds.
\item[\rm (c)] The measure\/ $\gamma^{}_{\sigma,m}$ does not depend on\/
   $\sigma\in C_c (G)$, provided\/ $\int_G \sigma\dd t =1$.
\item[\rm (d)] The measure\/ $\gamma^{}_{\sigma,m}$ is positive definite.
\end{itemize}
\end{lemma}
\begin{proof}
(a) We have to  show that  $\gamma^{}_{\sigma,m}$ restricted  to $C_K
(G)$ is continuous  for every compact $K$ in  $G$. Choose $\psi \in
C_c (G)$ non-negative with $\psi \equiv 1$ on $\supp (\sigma) + K$. Thus,
$$ |\varphi (s -t) \sigma (t) | \leq \psi (s) \|\varphi\|_\infty |\sigma| (t)$$for all $\varphi \in C_K (G)$ and  $s,t\in G$.
For $\varphi \in C_K (G)$ we can then estimate
\begin{eqnarray*}
|\gamma^{}_{\sigma,m} (\varphi) | &\leq& \left| \int_\varOmega \int_G \int_G |\varphi (s -t) \sigma (t)| d |\omega| (t)  \, d|\omega| (s)\, d m (\omega)\right|\\
&\leq & \int_\varOmega \int_G  \int_G \psi (s) \|\varphi\|_\infty |\sigma (t)| \, d  |\omega| (t)    \, d  |\omega| (s)\, d m (\omega)\\
&=& \|\varphi\|_\infty \int_\varOmega f_{\widetilde{\psi}}
(|\omega|) f_{|\widetilde{\sigma}|} (|\omega|) d m (\omega),
\end{eqnarray*}
and the statement follows.

(b) This follows by a direct computation (see Proposition $6$ of \cite{BL} as well).

(c)
Fix $\varphi\in C_c (G)$. By $\alpha$-invariance of $m$, we find that the map $\sigma \mapsto \
\gamma^{}_{\sigma,m} (\varphi)$ is $\alpha$-invariant and hence a  multiple of Haar measure on $G$. This shows the claim.

(d) This is a direct consequence of (b).
\end{proof}

The preceding lemma allows us to associate to any square integrable
probability measure an autocorrelation and a diffraction measure. They
are defined next.

\begin{definition}
Let $m$ be an $\alpha$-invariant square integrable probability
measure\/ $m$ on\/ $\MM$. Then, the measure $\gamma_m :=
\gamma^{}_{\sigma,m}$ for $\sigma\in C_c (G)$ with\/ $\int_G
\sigma(t)\dd t = 1$ is called the autocorrelation.  As $\gamma_m$ is
positive definite, its Fourier transform $\gammahat$ exists and is a positive measure on $\widehat{G}$. This measure is called the diffraction measure of the dynamical system.
\end{definition}

As discussed in the introduction to this section the usual approach to autocorrelation proceeds by an averaging
procedure along (models of) the substance in question. In our
framework, the substances are modeled by measures. Thus, we will have
to average  measures.  This is discussed in the remainder of this section.
It will turn out that averaging is possible once ergodicity is known. This is a consequence of the validity of Birkhoffs ergodic theorem (see the appendix and in particular  Lemma \ref{vanH2} for further details as well).   We will have to exercise quite some care as the functions $f_\varphi$
are not bounded.

\begin{definition} \label{definitionvanHove}A sequence $(B_n)$ of compact subsets of $G$ is
called a \textit{van Hove sequence} if
\begin{equation*}
   \lim_{n\to \infty} \frac{|\partial^K B_n|}{|B_n|} \; = \; 0
\end{equation*}
for all compact $K\subset G$. Here, for compact $B,K$, the
``$K$-boundary''  $\partial^K B $  of $B$ is defined as
\begin{equation*}
   \partial^K B \; := \; \overline{((B + K)\setminus B)} \cup
    [(\overline{G\setminus B} - K) \cap B],
\end{equation*}
where the bar denotes the closure.
\end{definition}
As discussed in the appendix, in our setting  there exists a van Hove sequence $(B_n)$ such that for any compact $K\subset G$ and any $\alpha$-invariant ergodic probability measure $m$ on $\MM$ and any $f\in L^1(\MM,m)$
$$ (\sharp)\;\:\;\:\;\:\; \lim_{n\to\infty} \frac{1}{|B_n|} \int_{\partial^K B_n}   |f(\alpha_t
(\omega))| dt = 0 $$ holds for $m$-almost every
$\omega\in \Omega$. Without loss of generality we can assume that $B_n = - B_n$ for all $n$. Fix such a sequence for the rest of this section.

\begin{lemma} \label{L12}
Let $m$ be an   $\alpha$-invariant square integrable ergodic
probability measure on\/ $\MM$. Then for all $\phi, \psi \in
C_c(G)$ nonnegative, there exists a $C<\infty$ and a set $\MM'$ of full measure in $\MM$ such that for all $\omega\in \MM'$  we have:
$$\limsup_{n \to \infty} \frac { \left| \tilde{\omega}_{B_n}\right| * \left| \omega_{B_n} \right|
 (\psi  * \tilde{\phi)} }{\left| B_n \right|}
\leq C\,.$$
\end{lemma}
\begin{proof}
Let $K$ be a compact  subset of $G$ with  $K=-K$ and $\supp
(\psi)$, $\supp(\phi) \subset K$. As a product of two
$L^2$-functions the function $f_{\tilde{\psi}} \circ |\cdot |
f_{\widetilde{\phi}} \circ |\cdot | $ belongs to  $L^1$,  $(\sharp)$ implies
(see  Proposition \ref{vanH2} as well) that almost surely
$$\lim_{n \to \infty} \frac{ \int_{B_n +K} T^t [f_{\tilde{\psi}} \circ |\cdot | f_{\widetilde{\phi}} \circ |\cdot |] dt}
{ \left| B_n \right| } = \int_{\MM} f_{\tilde{\psi}} ( \left| \mu
\right|) f_{\widetilde{\phi}} ( \left| \mu \right|)  dm(\mu)  =:C <\infty.$$

For $v,s \in B_n$ the function $ t \mapsto \psi (-t +v) \phi
(-t + s)$ is zero outside $B_n +K$ and
 hence
$$\int_{B_n+K} \psi (-t+v)
\phi (-t-s)dt = \int_{G} \psi (-t+v) \phi (-t-s)dt = \psi
*\tilde{\phi} (s + v).$$

Moreover, since $\phi, \psi$ are non-negative a short calculation gives that
$$T^t f_{\widetilde{\phi}} (|\omega|) = \int_G \phi(-s-t) d |\widetilde{\omega}|(s) \, ; \, T^t f_{\widetilde{\psi}} (|\omega|) = \int_G \psi(v-t) d |\omega|(v) \,.$$
Thus,
\begin{eqnarray*}
C &=&\lim_{n \to \infty} \frac {\int_{B_n +K} T^tf_{\tilde{\psi}}
(\left|\omega \right|) T^tf_{\tilde{\phi}} (\left|\omega
\right|) dt }{\left| B_n \right|} \\
 &=& \lim_{n \to \infty} \frac
{\int_{B_n +K}\int_G \int_G \psi (-t +v) \phi (-t - s) d \left|
\omega \right|(v) d \left| \tilde{\omega}
\right|(s) dt }{\left| B_n \right|}\\
&\geq& \limsup_{n \to \infty} \frac { \int_{B_n}\int_{B_n}
\int_{B_n+K} \psi (-t+v) \phi (-t-s)dt d \left| \omega \right|(v)
d
\left| \tilde{\omega} \right|(s)  }{\left| B_n \right|}\\
&=& \limsup_{n \to \infty} \frac {\int_{B_n}\int_{B_n} \psi
*\tilde{\phi} (s + v) d \left|
\omega \right|(v)     d \left| \tilde{\omega} \right|(s)}{\left| B_n \right|}\\
&=& \limsup_{n \to \infty} \frac { \left|
\tilde{\omega}_{B_n}\right| * \left| \omega_{B_n} \right|
 (\psi  * \tilde{\phi}) }{\left| B_n \right|}
\end{eqnarray*}
and the proof is finished.
\end{proof}

\begin{lemma} \label{L11}
  Let $m$ be an   $\alpha$-invariant square integrable ergodic probability
measure on\/ $\MM$. Let \/ $\phi, \psi \in C_c (G)$ be given. Then
$$\lim_{n \to \infty}\frac{ \tilde{\omega}*\omega_{B_n} (\tilde{\phi}* \psi) - \tilde{\omega}_{B_n}*\omega_{B_n}
(\tilde{\phi}* \psi)}{\left| B_n \right|}=0 \,,$$ almost surely in
$\omega$.
\end{lemma}
\begin{proof}
Let $K$ be a compact subset of $G$ with  $K=-K$ and $\supp (\phi), \supp(\psi)
\subset K$. Then,
\begin{equation*}
 (\tilde{\omega}*\omega_{B_n} -
\tilde{\omega}_{B_n}*\omega_{B_n}) (\tilde{\phi}* \psi) =  \int_G
\int_G \int_G  \overline{\phi( - v - s)} \psi(r-v)  1_{B_n}(s) (1 -
1_{B_n} (r))   d \omega (s) d \tilde{\omega} (r) dv.
\end{equation*}
For the integrand not to vanish we need $s\in B_n$, $r\in G\setminus B_n$ and
$v \in (B_n +K) \cap [(G \backslash B_n)+K] \subset \partial^K
B_n.$
Hence,  we can estimate
\begin{eqnarray*}
\left| (\tilde{\omega}*\omega_{B_n}  -
\tilde{\omega}_{B_n}*\omega_{B_n}) (\tilde{\phi}* \psi)\right|
&\leq & \int_{\partial^K B_n} \int_G \int_G \left|
\overline{\phi(- v - s)}\right| \left|\psi(r-v) \right|  d
\left|\omega \right|(s) d \left|\tilde{\omega} \right|(r) dv \\
&= & \int_{\partial^K B_n} T^{-v} [ f_{\left| \phi \right|}(\left|
\omega \right|) f_{\left|{\psi} \right|}(\left|\omega
\right|)]dv.
\end{eqnarray*}
The proof follows now from $(\sharp)$.
\end{proof}

\begin{theorem} Assume that $G$ is second countable.
Let $m$ be an   $\alpha$-invariant square integrable ergodic
probability measure\/ $m$ on\/ $\MM$. Then, almost surely in
$\omega$
$$\lim_{n \to \infty} \frac{\tilde{\omega}_{B_n}
* \omega_{B_n} } {\left| B_n \right|} = \gamma^{}_{m} \,,$$
where the limit is taken in the vague topology.
\end{theorem}
\begin{proof} The proof proceeds in three steps.

\textit{Step 1}: Let $\phi , \psi \in C_c (G)$, $t \in G$ be given
and set $ Z_n := \tilde{\phi} * \psi *\tilde{\omega} *
\omega_{B_n} (t)$. Then, $\lim_{n\to \infty} |B_n|^{-1} Z_n =
\langle f_\phi, T^t f_\psi \rangle$.

\smallskip

\textit{Proof of Step 1.} Let a
compact set $K$ in $G$ with $K= -K , 0 \in K$ and $\supp(\psi)
\subset K$ be given.  We are going to show that $ |B_n|^{-1} Z_n$
is of the same size as $ |B_n|^{-1} \int_{B_n}
\overline{f_\phi(\alpha_{t-v}(\omega))}f_\psi(\alpha_{-v}(\omega))dv$,
which by Birkhoff's ergodic theorem converges to $\langle f_\phi,
T^t f_\psi\rangle$.

A direct calculation  (see Theorem $5$ in \cite{BL} as well) shows
$$Z_n - \int_{B_n} \overline{f_\phi(\alpha_{t-v}(\omega))}f_\psi(\alpha_{-v}(\omega))dv =
\int_G(\overline{f_\phi(\alpha_{t-v}(\omega)}) D(v) dv \,$$ with
$D(v) := \int_G \psi(v-s) (1_{B_n}(s) - 1_{B_n}(v)) d \omega(s)$.
Then $D(v)$ is supported on $\partial^K B_n$  and hence
$$\Delta(n) :=\left| Z_n - \int_{B_n}
\overline{f_\phi(\alpha_{t-v}(\omega))}f_\psi(\alpha_{-v}(\omega))dv
\right| \leq  \int_{\partial^K B_n} \left|
\overline{f_\phi(\alpha_{t-v}(\omega))} D(v) \right| dv.$$

Now, note that $\left| D(v) \right| \leq  \int_G \left| \psi(v-s)
\right| d \left| \omega \right|(s) =  f_{\left| \psi \right|}(
\alpha_{-v} \left| \omega \right|)$, thus

$$\Delta(n)\leq  \int_{\partial^K B_n }
\left|
\overline{f_{\phi
}(\alpha_{t-v}(\omega))}
f_{|\psi|} (\alpha_{-v} |\omega|)
\right| d v
\,.$$
Application of $(\sharp)$ now  completes  the proof.

\medskip

\textit{Step 2} Let $D$ be a countable subset of $C_c (G)$. Then,
there exists a set $\Omega$ in $\MM$ of full measure with $
\lim_{n \to \infty} \frac{\tilde{\omega}_{B_n} * \omega_{B_n}
(\tilde{\phi} * \psi )} {\left| B_n \right|}=  \gamma^{}_{m} (
\tilde{\phi} * \psi )$ for all $\phi, \psi$ in $D$ and $\omega\in
\Omega$.

\smallskip

\textit{Proof of Step 2}. This follows immediately from Step 1, (b) of  Lemma \ref{Existenz}  and  Lemma \ref{L11}.

\medskip

\textit{Step 3} There exists a set $\Omega$ in $\MM$ of full
measure with $ \lim_{n \to \infty} \frac{\tilde{\omega}_{B_n} *
\omega_{B_n} (\sigma  )} {\left| B_n \right|}=  \gamma^{}_{m}
(\sigma)$ for all $\sigma \in C_c (G)$.

\smallskip

\textit{Proof of Step 3}.

Since $G$ is $\sigma$ compact, we can find a sequence $K_j$ of compact sets so that $G=\cup_{j} K_j$ and $K_j \subset K^\circ_{j+1}$.
It follows that $G=\cup_{j} K_j^\circ$ and in particular that each compact subset $K \subset G$ is contained in some $K_j$.

As $G$ is second countable, there exists a countable dense subset $D_j$ in each $C_{K_j}(G)$.

By second countability again, there exists furthermore  an approximate unit given by a sequence (i.e. a sequence $(\delta_n)$
in $C_c (G)$ such that $\varphi *\delta_n$ converges to $\varphi$ with respect to $\|\cdot \|_\infty$ for all $\varphi \in C_c (G)$).
Moreover, we can pick this sequence so that there exists a fixed compact set $0 \in K=-K$ such that $\supp(\delta_n) \subset K \, , \forall n$.

Let

$$D := (\cup_j D_j) \cup \{ \delta_n | n \in \NN \} \,.$$
Then $D$ is countable.

Lets observe that for each $\sigma \in C_c(G)$, there exists $j$ so that $\supp(\sigma) \subset K_j$. Thus, for all
$\epsilon >0$, there exists some $\psi \in D_j$ and $\phi=\delta_n$ so that $\| \sigma - \phi * \tilde{\psi} \|_\infty \leq \epsilon$, and
$\supp(\sigma), \supp(\phi * \tilde{\psi}) \subset K_j+K$.

For each $j \in \NN$ we can chose nonnegative $\phi_j, \psi_j\in C_c (G)$, $j\in \NN$ such that $\phi_j *
\tilde{\psi_j} \geq 1$ on $K_j+K$.

 By Lemma \ref{L12}, for each $j$ there exists a constant $C_j \geq 0 $ and a subset $\Omega_j$ of full measure so that for all $\omega \in \Omega_j$
 we have
$$\lim_{n \to \infty} \frac { \left| \tilde{\omega}_{B_n}\right| * \left| \omega_{B_n}
\right| (\phi_j * \tilde{\psi}_j) }{\left| B_n \right|} \leq C_j
\,.$$

Let $\Omega'$ be the set of full measure given by $D$ in step 2. Then $\Omega:= \Omega' \cap (\cap_{j} \Omega_j)$ has full measure.

Let $\sigma \in C_c(G)$ and $\omega \in \Omega$. Then there exists an $j$ so that $\supp(\sigma) \subset K_j$.

Let
$$C:= \max \{C_j+1 , |\gamma_m| (\phi_j * \tilde{\psi}_j ), 1 \}.$$

Since $\omega \in \Omega_j$, there exists an $N_0$ so that for all $n > N_0$ we have:

$$ \frac { \left| \tilde{\omega}_{B_n}\right| * \left| \omega_{B_n}
\right| (\phi_j * \tilde{\psi}_j) }{\left| B_n \right|} \leq C_j +1 \leq C
\,.$$

Let $\epsilon >0$. Then there exists $\psi, \phi \in D$ so that
$$| \sigma - \phi * \tilde{\psi} |\leq \frac{\epsilon}{C} \phi_j
*\tilde{\psi}_j . $$

When combined with the definition of $C$, this
gives easily
$$\left| \gamma^{}_{m} ( \sigma -\phi * \tilde{\psi)}) \right| \leq
\epsilon \,\mbox{and}\;\;\left| \frac{\tilde{\omega}_{B_n} *
\omega_{B_n} (\sigma - \phi*\tilde{\psi} )}{\left| B_n \right|}\right|
\leq \varepsilon \, \forall{ n > N_0}.$$

Moreover, since $\omega \in \Omega'$, by Step 2 we have
$$\left| \frac{\tilde{\omega}_{B_n} * \omega_{B_n} (\phi*\tilde{\psi} )} {\left| B_n \right|} - \gamma^{}_{m} (  \phi*\tilde{\psi}  )
\right| \leq \epsilon \, \forall n > N_1 \,.$$

Hence, for all $n > N:= \max \{N_0,N_1\}$ we have

$$\left| \frac{\tilde{\omega}_{B_n} * \omega_{B_n} (\sigma )} {\left| B_n \right|} - \gamma^{}_{m} ( \sigma  ) \right| \leq 3\epsilon \,.$$

\end{proof}

\section{Dynamical systems and  pure point diffraction}\label{Dynamical}
In this section, we relate spectral theory of measure dynamical
systems to diffraction theory. We will assume that we are given an
$\alpha$-invariant square integrable probability  measure $m$ on $\MM$ such that the associated unitary representation $T_m$ is strongly continuous. The associated
autocorrelation will be denoted by  $\gamma=\gamma_m$.  We will then discuss the
relationship between $\gammahat_m$ and the spectrum of the unitary
representation $T_m$. Our main result shows that, given (D),  pure pointedness of
$\gammahat_m$ is equivalent to pure pointedness of $T_m$. This
generalizes the corresponding results of \cite{BL,Gouere-1, LMS-1}.

\begin{prop} \label{spectralmass}
      The equation\/ $\rho^{}_{f_\varphi} =
   |\widehat{\varphi}|^2 \, \widehat{\gamma_m}$ holds for every $\varphi \in
    C_c (G)$.
\end{prop}
\begin{proof} The proof can be given exactly as in \cite{BL}. We include it for completeness reasons.
By the very definition of $\rho_{f_\varphi}$ above, the (inverse)
Fourier transform (on $\widehat{G}$) of  $\rho^{}_{f_\varphi}$
is\/ $t\mapsto \langle f_\varphi, T^t f_\varphi\rangle$. By
Lemma~\ref{Existenz}, we have $\langle f_\varphi , T^t
f_\varphi\rangle = \big(\widetilde{\varphi} \ast \varphi \ast
\gamma^{}_{m}\big) (t)$. Thus, taking the Fourier transform (on
$G$), we infer $\rho^{}_{f_\varphi} = |\widehat{\varphi}|^2 \,
\widehat{\gamma_m}$.
\end{proof}

Note that the closed $T$-invariant subspace $\cU$ of
$L^2(\MM,m)$ gives rise to a representation $T_{\cU}$ of $G$ on
$\cU$ by restricting the representation $T$ to $\cU$. The spectral
family of $T_{\cU}$ will be denoted by $E_{T_{\cU}}$.

\begin{definition}
  Let\/ $\rho$ be a non-negative measure on\/ $\widehat{G}$ and let\/ $S$ be an arbitrary
  strongly continuous unitary representation of\/ $G$ on an Hilbert space.
  Then, $\rho$ is called a {\em spectral measure}\/
  for\/ $S$ if the following holds for all Borel sets\/ $B$:
  $E_S(B)=0$ if and only if\/ $\rho(B)=0$.
\end{definition}

\begin{theorem}\label{Spectralmeasure} Let $m$ be a square integrable probability measure on $\MM$ with associated autocorrelation $\gamma=\gamma_m$. Then,
the measure\/ $\widehat{\gamma}$ is a spectral measure for\/ $T_\cU$.
\end{theorem}
\begin{proof} Given the previous results the proof follows as in
  \cite{BL}. We only sketch the details: Let $B$ be a Borel set in $\widehat{G}$. Then,
$E_{T_{\cU}} (B)= 0$ if and only if
$
   \langle f_\varphi, E_T (B) f_\varphi\rangle \; =\; 0$ for every $\varphi\in
   C_c (G)$.
By Proposition \ref{spectralmass}, we have $\rho^{}_{f_\varphi} =
|\widehat{\varphi}|^2 \widehat{\gamma_m}$ and,
in particular,
\begin{equation*}
   \langle f_\varphi, E_T (B) f_\varphi\rangle
   \; = \rho_{f_\varphi} (B) \; =  \int_B
     |\widehat{\varphi}|^2\dd \widehat{\gamma_m} \, .
\end{equation*}
These considerations show that $E_{T_{\cU}} (B)= 0$ if and only if
$0= \int_B |\widehat{\varphi}|^2\dd \widehat{\gamma_m}$ for every
function
$\varphi \in C_c (G)$.  By density, this is equivalent to $\widehat{\gamma_m}
(B) = 0$.
\end{proof}

The preceding considerations allow us to characterize the eigenvalues
 of $T_\cU$.  In this context, this type of result seems
to be new. It may be useful in other situations as well.

\begin{coro} Let $m$ be a square integrable probability measure on $\MM$ with associated autocorrelation $\gamma=\gamma_m$.
For $\varphi \in C_c (G)$ and $\lambda\in \Ghat$, the following
  assertions are equivalent:

\begin{itemize}

\item[(i)] $|\widehat{\varphi}|^2(\lambda)\gammahat(\{\lambda\}) >0$.

\item[(ii)] $E(\{\lambda\}) f_\varphi\neq 0$.

\item[(iii)] There exists an $f\neq 0$ with $f=E(\{\lambda\}) f$  in the closed convex  hull of $\{
  \overline{(\lambda,t)} T^t  f_\varphi : t\in G\}$.
\end{itemize}
\end{coro}
\begin{proof} Proposition \ref{spectralmass} gives
$$  \langle E(\{\lambda\}) f_\varphi,
E(\{\lambda\}) f_\varphi\rangle =  \rho_{f_\varphi} (\{\lambda\}) =
|\widehat{\varphi}|^2(\lambda)\gammahat(\{\lambda\})$$
and the equivalence between (i) and (ii) follows.  The implication
(iii)$\Longrightarrow$ (ii) is immediate as $E(\{\lambda\}) T^t f_\varphi =
(\lambda,t) E(\{\lambda\}) f_\varphi$ for every $t\in G$. It remains to show
(ii) $\Longrightarrow $ (iii).  Let $(B_n)$ be a van Hove sequence in
$G$.

As $\varphi \mapsto \frac{1}{|B_n|} \int_{B_n} \varphi (s) ds$ is a
probability measure on $G$,
the standard theory of vector valued integration (see e.g.  Chapter 3 in \cite{Rud2})  shows that the $L^2$-valued integral
$$ \frac{1}{|B_n|} \int_{B_n}
\overline{(\lambda,t)} T^t f_\varphi d t$$
belongs to the closed convex hull of $\{\overline{(\lambda,t)} T^t  f_\varphi : t\in G\}$
for every $n\in \NN$.
As von Neumann's ergodic theorem (see \cite{Kre})  gives
$$ E(\{\lambda\}) f_\varphi = \lim_{n\to \infty} \frac{1}{|B_n|} \int_{B_n}
\overline{(\lambda,t)} T^t f_\varphi d t,$$
where the limit is in the $L^2$-sense, the claim follows.
\end{proof}

Our main result reads as follows:

\begin{theorem} \label{main}  Let $m$ be an $\alpha$-invariant square integrable probability  measure on $\MM$  satisfying (D) with associated autocorrelation $\gamma=\gamma_m$.
The following assertions are equivalent:
\begin{itemize}
\item[\rm (i)] The measure $\widehat{\gamma}$ is  pure point.
\item[\rm (ii)] $T$ has pure point spectrum.
\end{itemize}
In this case, the group  generated by $\{\lambda \in \Ghat : \widehat{\gamma}(\{\lambda\}) >0\}$  is the set  of
eigenvalues of\/ $T$.
\end{theorem}
\begin{proof}
The implication (ii) $\Longrightarrow $ (i) is immediate from
  Theorem~\ref{Spectralmeasure}.

As for $(i)\Longrightarrow (ii)$, we note that $f_\varphi$ belongs to $ \cHp$ for every $\varphi\in C_c (G)$ by  Proposition~\ref{spectralmass}. By Corollary~\ref{maineins}, this implies that $g\circ f_\varphi$
  belongs to $\cHp$ for every $g\in C_c (\CC)$. By Lemma \ref{multiplication}, products of functions of the form $g\circ f_\varphi$, $g\in C_c (\CC)$, $\varphi \in C_c (G)$, then belong to $\cHp$ as well. Now, $(ii)$ follows from  (D).

It remains to show the last statement: set $L:=\{\lambda \in \Ghat :
\widehat{\gamma}(\{\lambda\}) >0\}$ and denote the group generated by $L$ in
$\Ghat$ by $S$. By Theorem \ref{Spectralmeasure} every $\lambda\in L$ is an
eigenvalue of $T_\cU$ and hence of $T$. As the eigenvalues form a group, we
infer that $S$ is contained in the group of eigenvalues of $T$. Moreover, by
Proposition  \ref{spectralmass}, the spectral measure $\rho_{f_\varphi}$
is supported on  $S$ (and even on $L$) for every $\varphi \in C_c (G)$. Thus,
by Corollary \ref{maineins}, the spectral measure $\rho_f$ is
supported on $S$ for every $f$ of the form $f =g\circ f_\varphi$ with $g\in C_c (\CC)$ and $\varphi \in C_c (G)$. By Lemma \ref{multiplication} this then holds as well for finite products of such functions.
As finite products of  such
functions  are total by (D), we infer   that the spectral
measure of every $f$ is supported on $S$. Thus, the set of eigenvalues is
contained in $S$.
\end{proof}

\textbf{Remark.} The implication  (ii) $\Longrightarrow $ (i) in Theorem \ref{main} holds even if
$m$ doesn't satisfy (D).

\section{Perturbation theory: Abstract setting} \label{Perturbation}
In this section, we shortly discuss a stability result for pure point
diffraction. In the topological setting, an analogous result is
discussed in \cite{BL2}. Our result is  more general in two ways:
First of all, the map $\varPhi$ below does not need to be continuous
but only  measurable. Secondly, the underlying space $\MM$ is
much bigger than the spaces considered in \cite{BL2} and hence we
obtain quite some additional freedom for perturbations.

\medskip

\begin{definition}
Let $m$ be an $\alpha$-invariant square integrable probability  measure on $\MM$. A measurable map
$\varPhi : \MM\longrightarrow \MM$ is said to satisfy condition (C) with
respect to $m$ if the following holds:

\begin{itemize}
\item $\varPhi \circ \alpha_t =\alpha_t \circ \varPhi$ for every $t\in G$.
\item The measure $\varPhi^\ast (m)$ defined by $\varPhi^\ast (m) (f):= m (f\circ \varPhi)$
  is square-integrable.
\end{itemize}
\end{definition}

If $\varPhi$ satisfies $(C)$ the measure $\varPhi^\ast (m)$ inherits
many properties of $m$. For example it can easily be seen to be ergodic if $m$ is
ergodic. Moreover, we have the following result on equivariant measurable perturbations.

\begin{theorem} \label{perturb}
Let $m$ be  an $\alpha$-invariant square integrable probability measure on $\MM$ satisfying (D)
  such that $\gammahat_m$ is a pure point measure supported on the
  group $S$. Let $\varPhi : \MM\longrightarrow \MM$ satisfy condition
  (C). Then, the dynamical system $(\MM,\varPhi^\ast (m))$ has pure
  point spectrum supported in $S$. In particular, the measure $
  \gammahat_{\varPhi^\ast (m)}$ is a pure point measure supported on
  $S$ as well.
\end{theorem}
\begin{proof} Set $n:=\varPhi^\ast (m)$. Denote the unitary representation of
  $G$ induced by $\alpha$ on $L^2 (\MM,m)$ and on $L^2 (\MM,n)$ by $T_m$ and
  $T_n$ respectively.  Note that $T_m$ is strongly continuous by assumption (D). Thus,  $T_n$ is strongly
 continuous as, by definition of $n$,  $\|T^t_n f - f\|_{L^2(n)} = \|T^t_m (f\circ \varPhi) - f\circ \varPhi\|_{L^2(m)}$ for all $f\in L^2 (\MM,n)$.
  The spectral measures associated to $f\in L^2 (\MM,m)
$ and $g\in L^2 (\MM,n)$ will be denoted by $\rho_f^m$ and $\rho_g^n$
  respectively. The definition of $n$ shows that
$$ \widehat{\rho_g^n } (t) = \langle g, T_n^t g\rangle = \langle g\circ
\varPhi, T_m^t g\circ \varPhi\rangle =  \widehat{\rho_{g\circ \varPhi}^m }
(t)$$
for all $t\in G$. This, gives
$$ \rho_g^n = \rho_{g\circ\varPhi}^m.$$ As $\gammahat_m$ is a pure
point measure supported on the group $S$, we infer from our main
result that $ \rho_{ f}^m$ is a pure point measure supported on $S$
for every $f\in L^2 (\MM,m)$. Thus, the preceding considerations show
that $\rho_g^n$ is a pure point measure supported on $S$ for every
$g\in L^2 (\MM,n)$. In particular, $T_n$ has pure point spectrum supported on $S$.  As
$ \gammahat_{n}$ is a spectral measure for a sub representation of
$T_n$ it must then be a pure point measure supported on $S$  as well.
\end{proof}

\bigskip

\subsection*{Acknowledgments} The authors are grateful to Michael Baake for various remarks which improved the manuscript.
D.L. would like to thank Jean-Baptiste Gou\'{e}r\'{e} for illuminating
discussions. N.S. would like to thank Robert Moody for stimulating
interest in this subject and NSERC for ongoing financial support.
Part of this work was done while one of the authors (N.S.) was
visiting TU Chemnitz. Financial support from DFG for this visit as
well as the hospitality of Fakult\"at f\"ur Mathematik is gratefully
acknowledged. Finally, the authors would like to thank the anonymous referee for a very careful reading of the manuscript, pointing out some inconsistencies  and   various  further well placed suggestions.

\newpage
\appendix
\section{Averaging
Sequences}\label{Appen}
In this appendix we consider the following situation. Let $X$ be a set
with a $\sigma$-algebra ${\mathcal B}$ and a measurable action $\alpha
: G\times X\longrightarrow X$ of the locally compact amenable group $G$ on
$X$. Let $\mu$ be an $\alpha$-invariant ergodic probability measure on
$X$.  As usual a sequence $(B_n)$ of compact subsets of $G$ is
called a \textit{F{\o}lner sequence} if
\begin{equation*} \frac{\lvert B_n \triangle
(B_n  K) \rvert }{ \lvert B_n \rvert} \xrightarrow   {n\to\infty} 0
\end{equation*}
for all compact $K\subset G$.  Here, $\triangle$ denotes the symmetric
difference.  We say that the Birkhoff ergodic theorem holds along the
F{\o}lner sequence $(B_n)$ if for any $f\in L^1 (X,\mu)$
$$\lim_{n\to \infty} \frac{1}{|B_n|} \int_{B_n} f(\alpha_t x) dt =
\int_X f(x) d\mu$$ for $\mu$-almost every $x\in X$. The aim of this
appendix is to show that any  F{\o}lner sequence  admits a subsequence $(B_n)$  such that  Birkhoff ergodic theorem holds along $(B_n K)$ for any compact
$K\subset G$ containing $0$. This will show that certain ``boundary terms'' which we meet in our considerations indeed go to zero.

\begin{definition}
A  F{\o}lner
 sequence $B_n$ is called \textit{tempered} if there
exists a constant $C>0$ so that
$\left| \cup_{ k < n} (B_k^{-1} B_n) \right| \leq C \left|B_n \right|$
\end{definition}

As shown by Lindenstrauss in \cite{Linden} the following holds:

\begin{itemize}

\item[(A)]  Every  F{\o}lner  sequence has a tempered subsequence.

\item[(B)] The Birkhoff ergodic theorem holds  along any tempered F{\o}lner sequence.

\end{itemize}

Here, (B) is one of the main results of \cite{Linden}.

\begin{lemma} \label{Lin} Let $(B_n' )_n$ be a  F{\o}lner sequence and
let $( K_l )_l$ be an arbitrary sequence of compact sets  in
$G$. Then, there exists a subsequence $(B_n)$ of $(B_n')$ so that
the Birkhoff ergodic theorem  holds simultaneously along  $( B_n  K_l
)_n$ for any $l\in \NN$.
\end{lemma}
\begin{proof}
Since $( B_n')$ is a F{\o}lner sequence, the
sequence $(B_n' K_l)_n$ is also a F{\o}lner sequence for any fixed $l$. Hence any subsequence of it is F{\o}lner again.   By $(A)$, we can then find a
subsequence $(B_{k(1,n)} )_n$ so that $(B_{k(1,n)}  K_1 )_n$ is
tempered. An inductive argument shows that for each $l$ there exists a
subsequence $( B_{k(l,n)} )_n$ of $( B_{k(l-1,n)} )_n$ so that
$(B_{k(l,n)} K_l )_n$ is tempered.  Then, by (B), Birkhoff's ergodic
theorem holds simultaneously along all $( B_{k(l,n)} K_l )_n$. A
simple diagonalization procedure now completes the proof.
\end{proof}


\begin{lemma}\label{vanH2}
Any F{\o}lner sequence  contains a
subsequence $(B_n)_n$ so that for   any compact  $K \subset G$ containing $0$ and any  $f \in L^1(\mu)$ we have
$$ \lim_{n \to \infty} \frac{\int_{B_n K} f(\alpha_t (x))
dt}{\left| B_n \right|}= \int_X f(y) d \mu(y)
$$
for $\mu$-almost every $x\in X$.
\end{lemma}
\begin{proof}

Since $G$ is $\sigma$-compact, we can find an increasing sequence of
compact sets $K_l$, $l\in \NN$ so that $K_1$ contains $0$, $K_l$ is
contained in the interior of $K_{l+1}$ for each $l\in \NN$ and the
union over all $K_l$ is just $G$.  Then, any compact $K \subset G$ is
a subset of some $K_l$. Now, let $( B_n )_n$ be the subsequence
defined by Lemma \ref{Lin}.  Let $f \in L^1(\mu)$ and a compact $K
\subset G$ with $0\in K$ be given and choose  $l$ with $K
\subset K_l$. Then,
\begin{eqnarray*} \left| \frac{\int_{B_n K} f(\alpha_t (x))
 dt   - \int_{B_n} f(\alpha_t (x))
 dt}{\left|B_n \right|} \right| &\leq&  \frac{\int_{(B_n K) \backslash
 B_n } \left| f(\alpha_t (x))\right| dt}{\left|B_n \right|} \leq
 \frac{\int_{(B_n K_l) \backslash B_n } \left|
 f(\alpha_t(\omega))\right| dt}{\left|B_n \right|}\\ &= &
 \frac{1}{|B_n|} \left( \int_{B_n K_l} |f(\alpha_t(x))| dt -
 \int_{B_n} |f(\alpha_t(x))| dt\right).
\end{eqnarray*}
 As $ \left|B_n K_l \right| / \left|B_n \right| \to 1$ when $n\to \infty$,  and Birkhoff's ergodic theorem holds along both $(B_n)$ and $(B_n K)$
the result follows easily.
\end{proof}

We now come to the desired result on the vanishing of boundary type terms.

\begin{prop}\label{vanH}
Let $( B_n )_n$ be a F{\o}lner sequence as in Lemma
\ref{vanH2}.  Then, for all $f \in L^1(\mu)$ and all compacts $K
\subset G$
$$\lim_{n \to \infty} \frac{\int_{ C_n } \left| f(\alpha_t (x))
\right| dt}{\left| B_n \right|} = 0$$
for $\mu$-almost every $x\in X$ along any sequence $(C_n)$ with  $C_n \subset B_n  K$ for all $n\in \NN$ and $|C_n|/|B_n|\longrightarrow 0$, $n\to \infty$.
\end{prop}
\begin{proof}
Let $\epsilon >0$ be arbitrary. Set $\widetilde{K}:=K\cup \{0\}$.

For $N \in \NN$ we define the function $f^N$ on $X$ by $ f^N (x):=
f(x)$ if $|f(x)| \leq N$ and $f(x) = 0$ otherwise. Then $\lim_{ N \to
\infty} f^N = f$ in $L^1 (X,\mu)$.  Therefore, there exists an $N\in
\NN$ with $\| f - f^N\|_1 \leq \epsilon$. By Lemma \ref{vanH2}, for
almost every $x\in X$, there exists an $n_1 = n_1 (x,\epsilon) $ so that for
all $n\geq n_1$, have
$$\int_{B_n K} |f (\alpha_t (x)) - f^N (\alpha_t (x))| dt \leq  \int_{B_n \widetilde{K}} |f (\alpha_t (x)) - f^N (\alpha_t (x))| dt \leq
2 \epsilon \left| B_n \right|.$$
Thus, for such $x$ and  $n \geq n_1$,
\begin{eqnarray*}
 \frac{\int_{C_n } \left| f(\alpha_t (x))
\right| dt}{\left| B_n \right|} &\leq& \frac{\int_{C_n} \left| f^N(\alpha_t (x)) \right| dt}{\left| B_n
\right|}+\frac{\int_{C_n} \left| (f(\alpha_t (x))-f^N(\alpha_t (x)))
\right| dt}{\left| B_n \right|} \\
&\leq & \frac{\int_{C_n} N dt}{\left| B_n
\right|}+\frac{\int_{B_n K} \left| (f(\alpha_t (x))-f^N(\alpha_t (x)))
\right| dt}{\left| B_n \right|} \\
&\leq & \frac{N \left| C_n \right|}{\left| B_n
\right|}+2\epsilon \,.
\end{eqnarray*}
As $|C_n|/|B_n|\to 0$ by assumption, we obtain
$$\frac{\int_{C_n } \left| f(\alpha_t (x)) \right| dt}{\left| B_n
\right|}\leq 3 \epsilon$$ for such $x$ and all large enough $n$. As
$\epsilon >0$ is arbitrary, the statement follows.  \end{proof}

When dealing with $\sigma$-compact, locally compact Abelian groups we
can do better than F{\o}lner sequences. Namely, in this case, there
exists a van Hove sequence as shown in \cite[p.~249]{Martin2}. Of
course, every van Hove sequence is a F{\o}lner sequence. In this case,
we can apply the previous Proposition with $C_n = \partial^K B_n
\subset B_n \widetilde{K}$, $n\in \NN$, and
$\widetilde{K}:=K\cup\{0\}$ compact.  This is used in Section
\ref{Diffraction}.

\end{document}